\documentclass[article,twocolumn]{revtex4-1}

\usepackage[english]{babel}
\usepackage{bm}
\usepackage{times}
\usepackage{ulem}
\usepackage{hyperref}
\usepackage[compat=1.0.0]{tikz-feynman}

\hypersetup{
    colorlinks=true,
    linkcolor=blue,
    filecolor=magenta,      
    urlcolor=blue,
    citecolor=blue,
    pdftitle={Sharelatex Example},
    pdfpagemode=FullScreen
    }
\usepackage{listings}
\usepackage{slashed}
\usepackage{color}
\usepackage{appendix}
\usepackage{mathtools}
\usepackage{epsfig}
\usepackage{dcolumn}
\usepackage{textcomp}
\usepackage{appendix} 
\usepackage{multirow}
\usepackage{graphicx}
\usepackage{soul}
\usepackage{subfigure}

\graphicspath{{PLots/}}

\newcommand{\Yoxara}[1]{{\color{magenta}#1}}

\begin{document} 

\begin{flushright}
MS-TP-23-20
\end{flushright}


\title{Searching for a Leptophilic $Z^\prime$ and a 3-3-1 symmetry at CLIC }

\author{A.Alves$^{1}$}
\email{aalves@unifesp.br}
\author{G. Gil da Silveira$^{9}$}
\email{gustavo.silveira@ufrgs.br; gustavo.silveira@cern.ch}
\author{V. P. Gonçalves$^{6,7}$}
\email{barros@ufpel.edu.br}
\author{F. S. Queiroz$^{2,3,4}$}
\email{farinaldo.queiroz@ufrn.br}
\author{Y. M. Oviedo-Torres$^{3,8}$}
\email[Corresponding author: ]{ymot@estudantes.ufpb.br}
\author{J. Zamora-Saa$^{2,5}$}
\email{jilberto.zamora@unab.cl; jilberto.zamorasaa@cern.ch}

\affiliation{$^1$Departamento de Física, Universidade Federal de São Paulo, UNIFESP, Diadema, São Paulo, Brazil}
\affiliation{$^2$ Millennium Institute for Subatomic Physics at High-Energy Frontier (SAPHIR), Fernandez Concha 700, Santiago, Chile}
\affiliation{$^3$ International Institute of Physics, Universidade Federal do Rio Grande do Norte, Campus Universitario, Lagoa Nova, Natal-RN 59078-970, Brazil}
\affiliation{$^4$ Departamento de Física, Universidade Federal do Rio Grande do Norte, 59078-970, Natal, RN, Brasil}
\affiliation{$^5$Center for Theoretical and Experimental Particle Physics - CTEPP, Facultad de Ciencias Exactas, Universidad Andres Bello, Fernandez Concha 700, Santiago, Chile}
\affiliation{$^6$ Institut f\"ur Theoretische Physik, Westf\"alische Wilhelms-Universit\"at M\"unster,
Wilhelm-Klemm-Straße 9, D-48149 M\"unster, Germany}
\affiliation{$^7$Institute of Physics and Mathematics, Federal University of Pelotas, 
  Postal Code 354,  96010-900, Pelotas, RS, Brazil}
\affiliation{$^8$ Departamento de Física, Universidade Federal da Paraíba, Caixa Postal 5008, 58051-970, João Pessoa, PB, Brazil}
\affiliation{$^9$ Instituto de Física, Federal University of Rio Grande do Sul,
Av. Bento Gonçalves, 9500, CEP 91501-970, Porto Alegre, Brazil}

\begin{abstract}
We derive the discovery potential of a leptophilic $Z^\prime$, and a $Z^\prime$ rising from a $SU(3) \times SU(3)_L \times U(1)_N$ symmetry at the Compact Linear Collider (CLIC), which is planned to host $e^+e^-$ collisions with 3 TeV center-of-mass energy. We perform an optimized selection cut strategy on the transverse momentum, pseudorapidity, and invariant mass of the dileptons in order to enhance the collider sensitivity.
We find that CLIC can potentially reach a $5\sigma$ signal of a $1-3$~TeV leptophilic $Z^\prime$ with less than $1fb^{-1}$ of integrated luminosity. As for the $Z^\prime$ belonging to a 3-3-1 symmetry, CLIC will offer a complementary probe with the potential to impose $M_{Z^\prime} > 3$~TeV with $\mathcal{L}=2fb^{-1}$.
\noindent

\end{abstract}


\maketitle
\flushbottom

\section{\label{SectionI}Introduction}

The Compact Linear Collider (CLIC) represents a high-luminosity, TeV-scale linear ${e}^{+}{e}^{-}$ collider under vigorous development by global collaborations, centered at CERN \cite{Brunner:2022usy}. In order to maximize its physics potential, a staged construction is foreseen. CLIC will operate at center-of-mass energies of 380 GeV, 1.5 TeV, and 3 TeV. The total site length is projected to vary between 11 km and 50 km. Considerable advancements in recent years have been made, both in terms of technical development and system testing for the CLIC accelerator which have greatly reduced its building cost and improved its physics potential. The first beams are planned to occur by 2035, thereby initiating a physics program expected to span over 25 to 30 years. Giving our successful experience with the Large Electron Positron collider \cite{Kawamoto:2001lso, Rolandi:1995ng, ALEPH:2006bhb}, with its larger luminosity and center of mass
CLIC promises to be a unique opportunity to probe physics Beyond the Standard Model (BSM). It will facilitate direct searches and enable a comprehensive range of precision measurements of Standard Model processes. This potential will be particularly valuable for studying the Higgs boson and extended scalar sectors as well as new gauge bosons. \newline

Several theories extending beyond the Standard Model (SM) rely on the existence of neutral gauge bosons that feature coupling to leptons. These theories provide potential explanations for unresolved phenomena, such as dark matter (DM), neutrino masses, the muon anomalous magnetic moment, among others. Such neutral bosons are often associated with a new Abelian gauge symmetry which is spontaneously broken, generating a $Z^\prime$ boson with mass around the new physics scale. Non-Abelian gauge symmetries can also give rise to a $Z^\prime$ boson. Such a $Z^\prime$ boson can be produced at Hadron and Lepton colliders producing a high-mass dilepton resonance. To the experimentalist, a $Z^\prime$ boson is, at the end of the day, a resonance more massive than the SM Z boson. Conversely, for a theorist, a $Z^\prime$ field represents a new force carrier, which holds the secret to the road of physics beyond the Standard Model (SM). \newline

Motivated by the importance of $Z^\prime$ bosons in theoretical constructions, we will assess the CLIC potential to discover a leptophilic $Z^\prime$ using a simplified model, and a $Z^\prime$ rising in extended gauge sectors with a $SU(3)\times SU(3) \times U(1)_N$ symmetry, 3-3-1 for short.  A leptophilic $Z^\prime$ emerges in simple gauged lepton number theories \cite{Heeck:2011wj,delAguila:2014soa,Bell:2014tta,Kara:2014zfc,Patra:2016shz,Heeck:2016xkh,Altmannshofer:2016brv,Altmannshofer:2016jzy,Sadhukhan:2020etu,Anchordoqui:2021vrg,Asai:2021xtg,Moroi:2022qwz}, and or in more complex setups \cite{Ko:2010at,Kara:2011xw,Allanach:2015gkd,Buras:2021btx,Li:2022okq,Chun:2021rtk,Li:2023lin}. The $SU(3)_C \times SU(3)_L \times U(1)_N$ symmetry has featured in a multitude of papers in the literature because it could help to solve neutrino masses \cite{Tully:2000kk, Montero:2001ts, Cortez:2005cp, Cogollo:2009yi,Cogollo:2010jw, Cogollo:2008zc, Okada:2015bxa, Pires:2018kaj, CarcamoHernandez:2020pnh}, flavor puzzles \cite{Cabarcas:2012uf,Hue:2017lak}, dark matter \cite{Fregolente:2002nx,Hoang:2003vj,deS.Pires:2007gi,Mizukoshi:2010ky,Ruiz-Alvarez:2012nvg,Profumo:2013sca,Dong:2013ioa,Dong:2013wca,Cogollo:2014jia,Dong:2014wsa,Dong:2014esa,Kelso:2014qka,Dong:2014esa,Mambrini:2015sia,Dong:2015rka,deSPires:2016kkg,Alves:2016fqe,RodriguesdaSilva:2014gbi,Carvajal:2017gjj,Dong:2017zxo,Arcadi:2017xbo,Montero:2017yvy,Huong:2019vej,Alvarez-Salazar:2019cxw,VanLoi:2020xcq,Dutra:2021lto,Oliveira:2021gcw}, among others. \newline


In the context of linear colliders, there have been leptophilic $Z^\prime$ studies in the past \cite{Soa:2003fr,Blaising:2012tz,delAguila:2014soa,Blaising:2012tz,Spor:2022zob,Inan:2021pbx,Chun:2021rtk,Li:2023lin,Gurkanli:2023ktk}. Although, none of them were focused on a sequential leptophilic $Z^\prime$ boson, which has couplings to SM lepton equal to the Z-boson, with no interaction with quarks. A sequential $Z^\prime$ is often the target of several collider searches at the LHC. If couplings to quarks are turned off, a sequential leptophilic $Z^\prime$ arises and could be a benchmark model at CLIC. Obviously, in the presence of couplings to quarks, the LHC reach is significantly better than CLIC. However, in the absence of interactions with quarks, CLIC stands out as a promising probe. Moreover, CLIC can also serve as a precision machine once a $Z^\prime$ boson is discovered at LHC or at the High-Luminosity LHC.
It is important to emphasize that, using LHC data, collider bounds on the $Z^\prime$ bosons belonging to a 3-3-1 symmetry have been derived in the past, excluding $Z^\prime$ masses below 4 TeV. However, if the $Z^\prime$ is leptophilic, the relevant collider constraints mainly come from LEP, and masses larger than 200 GeV are not excluded, which motivates the searching at future $e^+e^-$ colliders. We have no data at the moment, thus we can only assess CLIC sensitivity reach. We will see that with less than $1fb^{-1}$ of luminosity, CLIC can discover such a $Z^\prime$ boson. Therefore, in spite of not being able to produce on-shell $Z^\prime$ bosons with masses larger than $3$~TeV, CLIC will clearly be able to better constrain the couplings and properties of any new gauge boson that has sizeable couplings to leptons. In this spirit, we will perform a phenomenological analysis based on optimized cuts that yield better signal efficiency for both discovery and exclusion for a given luminosity in the $e^+e^- \to e^+e^-$ channel. Although our focus is on the impact of a leptophilic $Z^{\prime}$ in the $e^+e^- \to e^+e^-$ process, we will also present, for comparison and completeness of our analysis, the predictions from the 3-3-1 model. \newline

This paper is organized as follows. In the next Section we will describe the details of our analysis. In particular, we  briefly present the CLIC experiment and the models considered. Moreover,  the Monte Carlo simulation and cuts assumed in the analysis will be discussed. In Section \ref{SectionIII} we present our results for the total and differential cross sections derived considering a  leptophilic $Z^{\prime}$. The predictions associated to the 3-3-1 model are also presented for comparison.  The best kinematic cuts are obtained and the luminosity required to exclude (or discover) a $Z^{\prime}$ at CLIC is estimated considering distinct values for the $Z^{\prime}$ mass. Finally, in Section \ref{SectionIV}, we summarize our main conclusions.

\section{\label{SectionII}Details of the analysis}

\begin{figure}[t]
    \centering
     \includegraphics[scale=0.15]{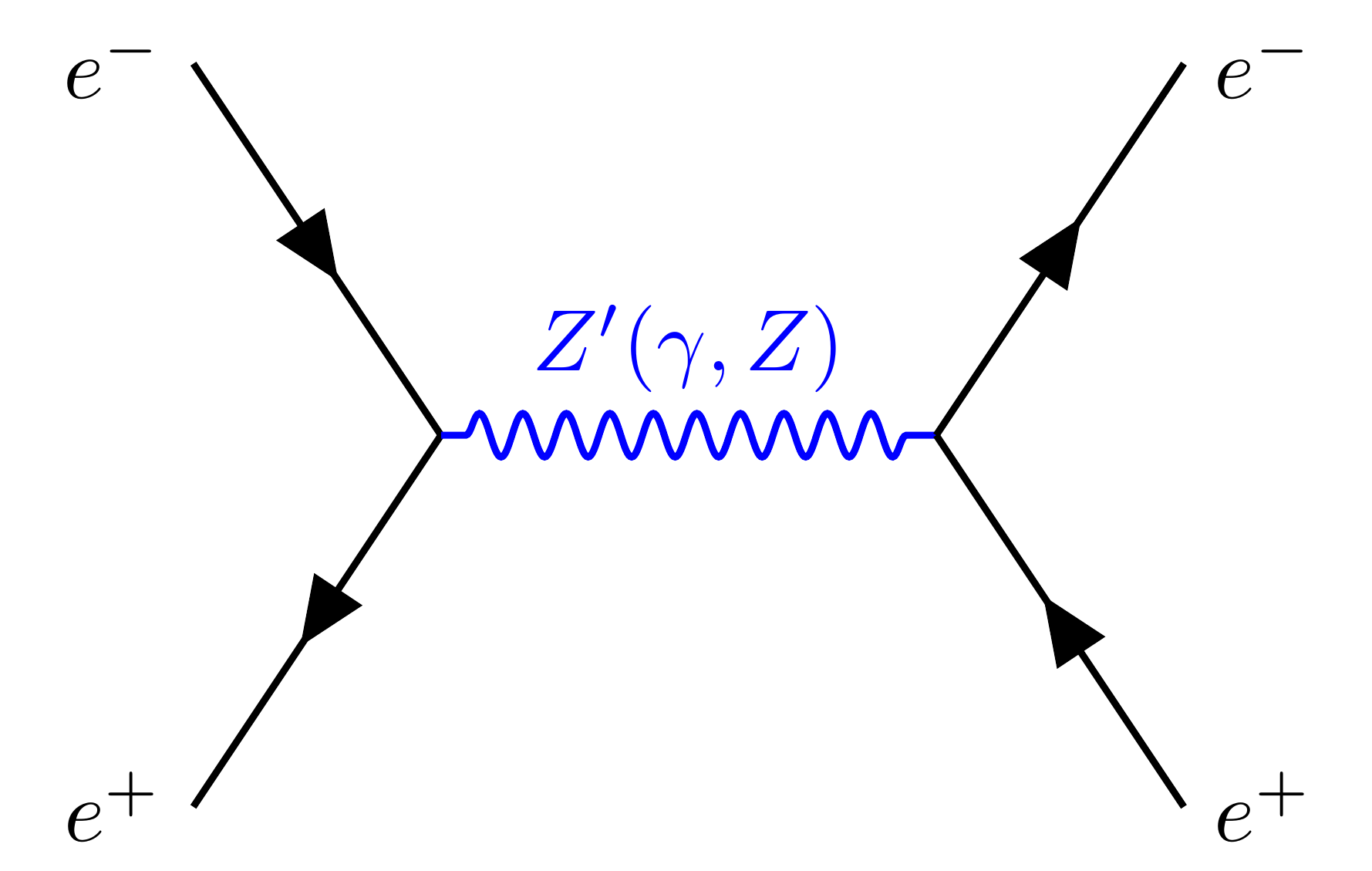}
    \includegraphics[scale=0.125]{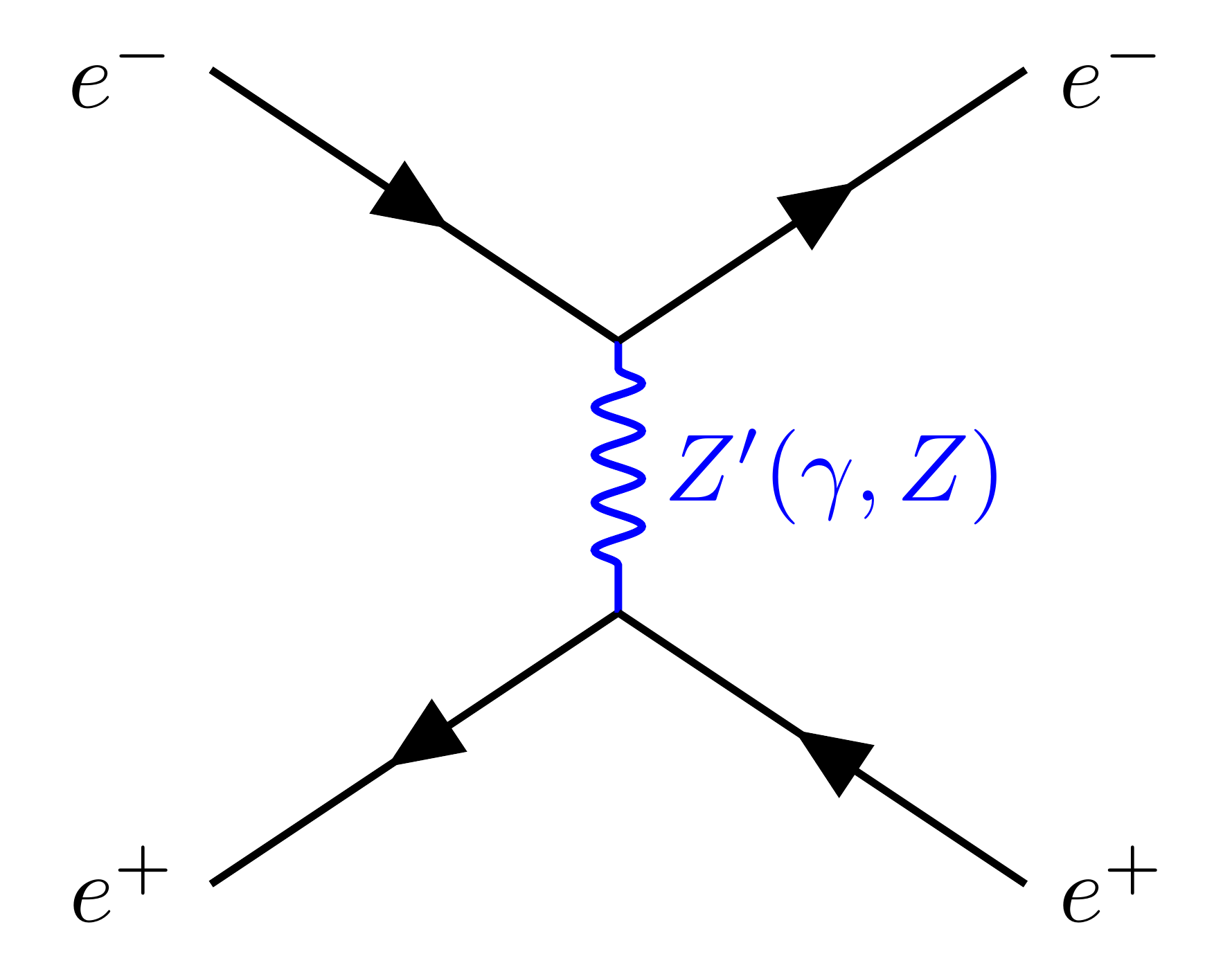}
    \caption{Feynman diagrams for the signal and background processes relevant for the $Z^\prime$ search at the CLIC.}
\label{signal_background_feynman_diagram}
\end{figure}

Our goal in present analysis is to estimate the impact of a $Z^{\prime}$  on the  $e^+ e^- \to e^+ e^-$ process. Such a boson can contribute for the direct electron - positron annihilation, represented in the left panel of Fig. \ref{signal_background_feynman_diagram}, as well can be present in the $t$-channel diagram (right panel). The Standard Model backgrounds are represented by these same diagrams but with the exchange of a photon or a $Z$ boson. We will focus on $e^+e^-$ collisions at CLIC and will assume a leptophilic $Z^{\prime}$.
Having in mind the new physics potential of CLIC and the reasonable hypothesis of ${Z}^{\prime}$ fields in theoretical constructions, in what follows we will briefly discuss the Compact Linear Collider and describe the models that we investigate in this work. \newline

\subsection{ The Compact Linear Collider}

The Compact Linear Collider (CLIC), currently under the development auspices of the CLIC accelerator collaboration, is a high-luminosity linear collider capable of reaching multi-TeV energies. The Conceptual Design Report (CDR) for the CLIC was released in 2012 \cite{Aicheler:1500095}. A pioneering feature of the CLIC is its utilization of the two-beam acceleration technique, which employs novel accelerating structures functioning within the range of 80-100~MV/m. \newline

The primary objective of this report was to substantiate the viability of the CLIC accelerator at elevated energies of 3~TeV. Furthermore, it was crucial to affirm that the presence of particles from beam-induced backgrounds and the characteristics of the luminosity spectrum would not inhibit the performance of high-precision physics measurements \cite{Aslanides:2706890,Brunner:2022usy,Linssen:2012hp}. In the same way that LEP and SLAC were important to test various SM predictions, the CLIC hopes to collect more stringent electroweak precision measurements and signals of new physics for a period of 27~years in three different but complementary stages~\cite{Robson:2018zje}. Moreover, based on a novel acceleration system, CLIC plans to progressively reach energies of up to $\sqrt{s}=3$~TeV and an integrated luminosity of 5~ab$^{-1}$ \cite{Franceschini:2019zsg,Sicking:2020gjp,Zarnecki:2020ics,CLICPhysicsWorkingGroup:2004qvu} in a sequence of stages. For the first stage of the CLIC operation, beams are expected to be delivered at an energy of $\sqrt{s} = 380$~GeV and a luminosity of 1~ab$^{-1}$. In the second and third stages, operations are aimed at elevated energies with beams at $\sqrt{s} = 1.5$~TeV with $\mathcal{L}=2.5$~ab$^{-1}$, and $\sqrt{s} = 3$~TeV with  $\mathcal{L}=5$~ab$^{-1}$, respectively. In each of these operation stages, the CLIC program aims to improve electroweak precision measurements of SM parameters and find, either indirect or direct, indications of new physics (NP)~\cite{CLIC:2018fvx}. \newline

These stages of the CLIC operation will be adjusted based on the results obtained by the HL-LHC. In particular, the center-of-mass energy can be increased by extending the length of the accelerator or by improving acceleration technology. In this way, the novel acceleration technology of the CLIC opens the possibility of carrying out studies with high center-of-mass energy and luminosity, accessing production mechanics at far higher energy than those observed at LEP. \newline

\subsection{The Models}

\subsubsection{Leptophilic $Z^\prime$}

A sequential $Z^\prime$ is a benchmark model at the LHC. It refers to the case where the $Z^\prime$ field couples to SM fermions in the same way the Z boson does \cite{CMS:2014lcz,ATLAS:2017fih}. Having that in mind, we propose to search for a leptophilic $Z^\prime$ with sequential couplings to SM leptons. In other words, it is a $Z^\prime$ that features only couplings to SM leptons identical to the Z boson. Therefore, the neutral current reads, 

\begin{equation}
\mathcal{L}=\frac{-g}{2\cos\theta_W}\sum_i \bar{\psi_i} \gamma^\mu (g_V^i-\gamma_5 g_A^i)\psi_i Z_\mu   
\label{eq:L-leptophilic}
\end{equation}

where $\theta_W=\tan^-1(g^\prime/g)$, $g_V^i= \tau_{3L}(i) 2 Q_i \sin^2\theta_W$, and $g_A^i=\tau_{3L}$. In this case, i=1,2,..6;  with $\psi_i$ running over all six lepton flavors. This new boson can potentially couple new particles but we hypothesize that these new particles are too heavy to be produced at CLIC via $Z^\prime$ exchange. The lagrangian above is, therefore, the relevant one for phenomenological studies. \newline


\subsubsection{3-3-1 Model}

The 3-3-1 model refers to the $SU(3)_C \times SU(3)_L \times U(1)_Y$ symmetry \cite{Pisano:1992bxx,Foot:1992rh,Foot:1994ym}. An obvious consequence of enlarging SM gauge group is the appearance of new vector bosons. Particularly, within the 331 models we have three new gauge bosons, six new fermions (three leptons and three quarks), and six new scalars. In these models, to avoid chiral anomalies and reproduce the correct SM interactions, leptons are accommodated in the triplet representations. In the quark sector, the first two generations transform under the triplet representation and the third generation transforms under the anti-triplet representation. Right-handed fermions transform in the singlet representation, as usual. Fermion masses are obtained through a two-step spontaneous symmetry breaking using three scalar triplets. There is freedom in the choice for the third component of the fermion triplet, and this freedom gives rise to different particle contents. We will focus on the 331RHN and 331LHN models~\cite{Alves:2022hcp}, where the third component of the lepton triplet is a right-handed neutrino or a heavy and neutral fermion, respectively. In both 331 RHN and LHN models, the lagrangian that describes the interaction between fermions and the Z' boson is the same lagrangian presented in Eq.~\eqref{eq:L-leptophilic}, where ${g}^{i}_{V}$ and ${g}^{i}_{A}$ are the vector and axial couplings whose values are shown in Table \ref{cc}. Again, we will tacitly assume that the new fermions and scalars of the 331 models are heavier than the $Z^\prime$ so the interactions above are the only ones relevant for a $Z^\prime$ search at CLIC. \newline

\begin{table}[t]
\begin{footnotesize}
\begin{center}
\begin{tabular}{|c|c|c|}
\hline
\multicolumn{3}{|c|}{$Z^{\prime}$ Interactions with the SM fermions in the 3-3-1 model} \\
\hline
Interaction &  $g^{i}_V$ & $g^{i}_A$   \\ 

\hline
$Z^{\prime}\ \bar u u,\bar c c  $ &
$\displaystyle{\frac{-3+8\sin^2\theta_W}{{6\sqrt{3-4\sin^2\theta_W}}}}$  & 
$\displaystyle{-\frac{1}{2\sqrt{3-4\sin^2\theta_W}}}$  \\
\hline
$Z^{\prime}\ \bar t t$ & 
$\displaystyle{\frac{-3-2\sin^2\theta_W}{{6\sqrt{3-4\sin^2\theta_W}}}}$  & 
$\displaystyle{-\frac{1-2\sin^2\theta_W}{2\sqrt{3-4\sin^2\theta_W}}}$  \\
\hline
$Z^{\prime}\ \bar d d,\bar s s  $ &
$\displaystyle{\frac{-3+2\sin^2\theta_W}{6\sqrt{3-4\sin^2\theta_W}}}$  & 
$\displaystyle{-\frac{{3-6\sin^2\theta_W}}{6\sqrt{3-4\sin^2\theta_W}}}$  \\
\hline
$Z^{\prime}\ \bar b b$ & 
$\displaystyle{\frac{-3+4\sin^2\theta_W}{{6\sqrt{3-4\sin^2\theta_W}}}}$  & 
$\displaystyle{-\frac{1}{2\sqrt{3-4\sin^2\theta_W}}}$  \\
\hline
$Z^{\prime}\ \bar \ell \ell $ &
$\displaystyle{\frac{1-4\sin^2\theta_W}{2\sqrt{3-4\sin^2\theta_W}}}$ &
$\displaystyle{\frac{1}{2\sqrt{3-4\sin^2\theta_W}}}$ \\
\hline
$Z^{\prime}\ \overline{\nu_{\ell}} \nu_{\ell} $ &
$\displaystyle{\frac{-\sqrt{3-4\sin^2\theta_W}}{18}}$ &
$\displaystyle{-\frac{\sqrt{3-4\sin^2\theta_W}}{18}}$ \\ 
\hline
\end{tabular}
\end{center}
\end{footnotesize}
\caption{Vector and Axial couplings of the ${Z^{\prime}}$ boson with fermions in the 331RHN and 331LHN models.}
\label{cc}
\end{table}


\subsection{Simulation}

All signal and background events were simulated using \texttt{FeynRules}~\cite{Feynrules}, \texttt{MadGraph5}~\cite{alwall2014automated}, \texttt{Pythia8}~\cite{Sjostrand:2007gs}, and  \texttt{Delphes3}~\cite{deFavereau:2013fsa}. In our analysis, we generate, in each case, 80k signal events for five $Z^\prime$ masses: 0.5, 1, 1.5, 2, and 2.5 TeV, and 200k for backgrounds. \newline 

We simulate ${e}^{+}{e}^{-}$ collisions at CLIC employing the \texttt{clic3000ll} Parton Distribution Function (PDF) set which takes initial state radiation effects at Leading Order+Leading Log approximation and also a parametrized implementation of beamstrahlung~\cite{Frixione:2021zdp}.  In this study, we assume that the new particle spectrum is heavy 
such that the decay of the ${Z}^{\prime}$ into BSM particles is kinematically forbidden. The interactions that we are considering are those corresponding to the ${Z}^{\prime}$ to SM fermions where the couplings to leptons are equal to the SM Z boson, but $Z^\prime$ does not couple to quarks, in the case of the leptophilic new gauge boson, as defined in Eq.~\eqref{eq:L-leptophilic}. In the case of 331 models, couplings to quarks diminish the branching fraction of the $Z^\prime$ into SM leptons and increase the total width compared to the leptophilic case for the same $M_{Z^\prime}$. In both models, we expect to observe sharp peaks in the $e^+e^-$ invariant mass but the leptophilic model will produce narrower resonances. \newline

The SM background and the signal have $s$ and $t$-channel diagrams exchanging a $Z,\gamma$ and $Z^\prime$, respectively, as depicted in panels (a) and (b) of Figure~\ref{signal_background_feynman_diagram}. To avoid collinear divergences,  we selected events that satisfy the following basic selection requirements
\begin{equation}
    p_T(l) > 100\; \hbox{GeV},\;\; |\eta(l)| < 3\; ,
    \label{basic_cuts}
\end{equation}
for both electron and positron in the event. \newline

\begin{figure}[!ht]
    \includegraphics[width=\columnwidth]{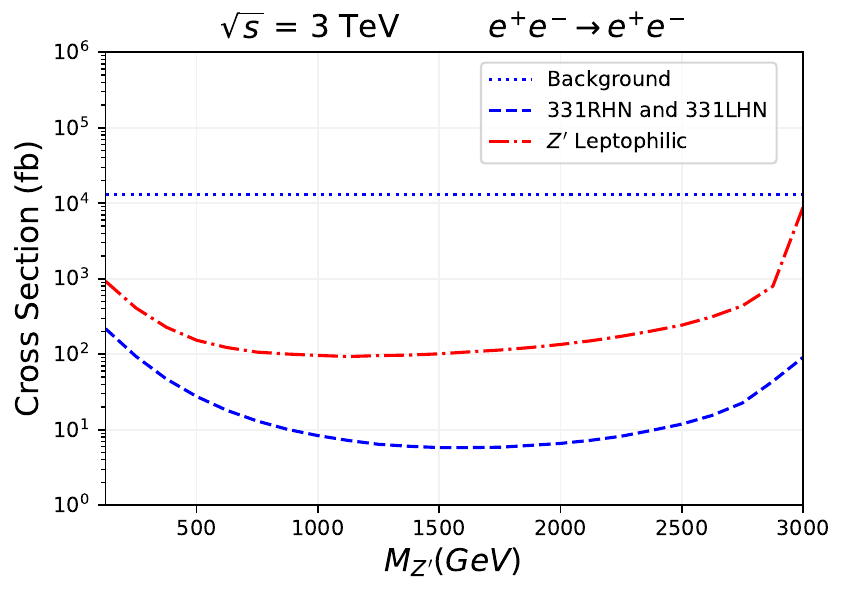}
    \caption{The $e^+ e^- \to e^+ e^-$ cross section for the leptophilic (red dashed lines) and 3-3-1 models (blue dashed lines). The SM background is displayed as a dotted blue curve. The basic cuts of Eq.~\eqref{basic_cuts} were required in all cases.}
\label{signalCS}
\end{figure}

\begin{figure*}[!ht]
    \centering
     \subfigure[]{\includegraphics[scale=0.5]{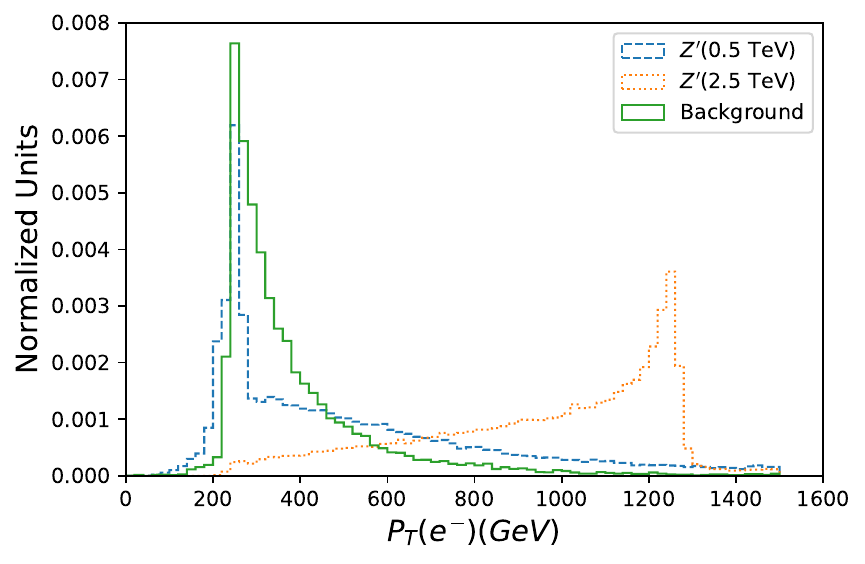}}
     \subfigure[]{\includegraphics[scale=0.5]{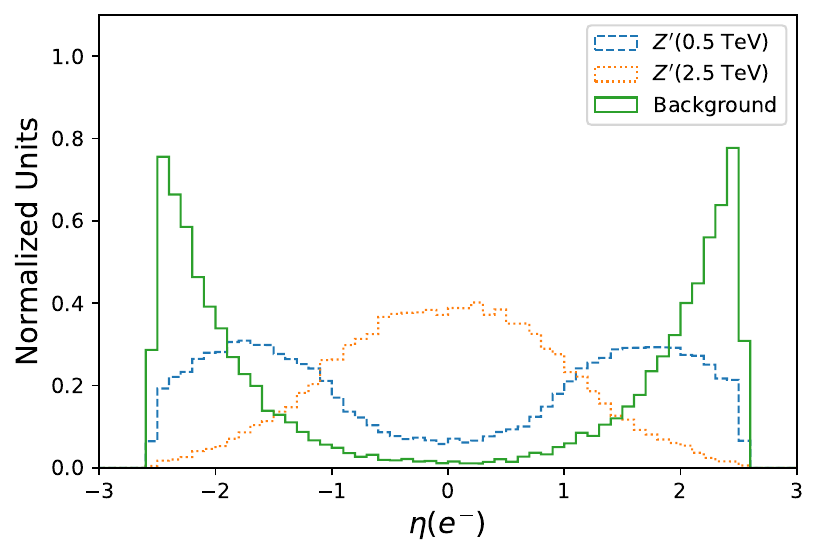}}
    \subfigure[]{\includegraphics[scale=0.5
]{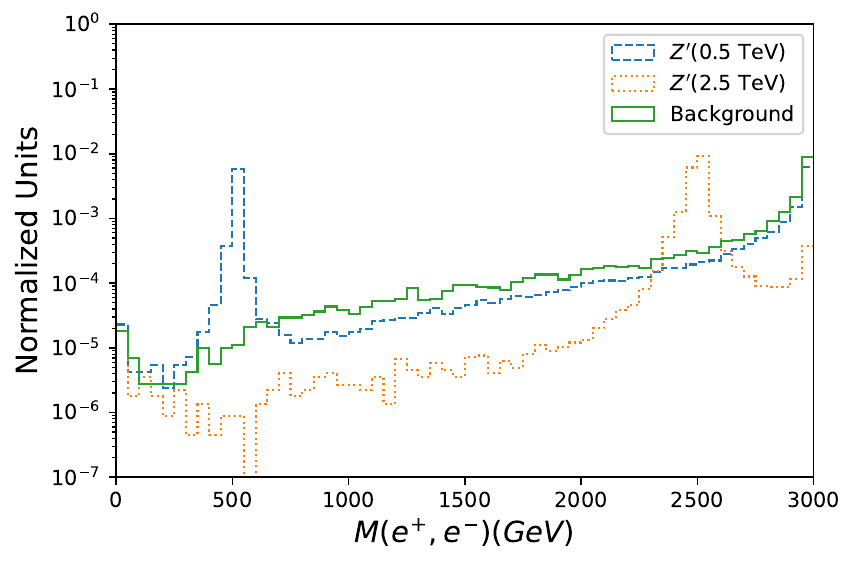}}
    \caption{The left(right) upper panel depicts the transverse momentum(rapidity) distribution of electrons of the SM background and signals corresponding to $Z^\prime$ masses of 0.5 and 2.5 TeV. In the lower panel, we show the $e^+e^-$ invariant masses distribution of background and signals. All signals correspond to leptophilic $Z^\prime$ model. The distributions of 331 signals are similar.}
\label{331LHN_all_BMsx}
\end{figure*}

\section{\label{SectionIII}Results and discussions}

In Figure~\ref{signalCS} we present our results for the $e^+e^- \to e^+e^-$ cross section as a function of the $Z^{\prime}$ mass, derived considering the cuts discussed in the previous section and assuming the Leptophilic and 3-3-1 models.  One has that 
for masses much smaller than the collider energy, the $t$-channel diagram dominates as the $Z^{\prime}$ is produced off its mass-shell, but heavier $Z^\prime$ bosons tend to be produced more in the mass-shell increasing the cross section towards $M_{Z^\prime}=3$ TeV. By their turn, 331 bosons present a smaller cross section once, in this case, there are more options for the $Z^\prime$ to decay into, contrary to the case of the leptophilic ones. In 331 and leptophilic models, $BR(Z^\prime\to e^+e^-) = 2.4$\%, and 11\%, respectively, for the masses range we are considering in this work. The SM cross section for the  $e^+e^-\to e^+e^-$ process is 13.2 pb and is represented by the blue dotted line in Fig.~\ref{signalCS} . \newline

The signal and background distributions of the electron transverse momentum, $p_T(e^-)$, electron rapidity, $\eta(e^-)$, and the $e^+e^-$ invariant mass, $M(e^+,e^-)$,  are shown in Figure~\ref{331LHN_all_BMsx}, for leptophilic $Z^\prime$ signals. The 331 $Z^\prime$ distributions  are similar. As we see, the signal and background features are very distinctive, especially for heavy $Z^\prime$ bosons. Nonetheless, for lighter $Z^\prime$, the peaks in $e^+e^-$ invariant mass still denounce the presence of signals making the distinction against the smooth background spectrum an easy task. Concerning the rapidity distributions, lighter $Z^\prime$ and the SM backgrounds display a similar behavior with the majority of events hitting the high rapidity regions of the detector, while heavy $Z^\prime$ produced  electrons and positrons at central rapidities. This is due to the competition between the $s$ and $t$-channel amplitudes, where the $t$-channel contribution is amplified when the final state lepton is collinear with the initial state one, while the $s$-channel produces high-$p_T$ yields. \newline 

To remove backgrounds and increase the statistical significance of the signal, we searched for the kinematic cuts on $p_T$, $\eta$, and $M(e^+,e^-)$ that maximize the signal efficiency and minimize the background one. As we discuss in the next section, the background efficiencies after optimization were found very small for $Z^\prime$ masses from 10 GeV to 3 TeV. We compute the signal significance as follows
\begin{equation}
    N_\sigma = 
     \frac{L\times\epsilon_S\sigma_S}{\sqrt{L\times\epsilon_B\sigma_B+(\varepsilon_B^{sys}\times L\times\epsilon_B\sigma_B)^2}},\; 
\end{equation}
where $\sigma_S$($\epsilon_S$), and $\sigma_B$($\epsilon_B$) represent the cross section (selection efficiency) of the signal and the backgrounds, respectively. The integrated luminosity is denoted by $L$, while $\varepsilon_B^{sys}$ is the systematic uncertainty in the background rate. In those cases where no Monte Carlo background events pass the selection requirements, we conservatively assume a background rate given by $\sigma_B/n_{MC}$, where $n_{MC}=2\times 10^5$, the number of simulated background events. \newline

\begin{table}[t]
\centering
\begin{tabular}{|c|c|c|c|c|} 
\hline
${M}_{{Z}^{\prime}}$ & ${p}_{T}(l)>$ & $|\eta(l)|<$ & $|M_{ee}-M_c| < \delta_M$ & $\epsilon_S$(\%) \\ 
\hline
   500 & 207 & 2.23 & $458 \pm 57$ & 1.9 \\ 
\hline
  1000  & 420 & 2.1 & $1041 \pm 51$ & 9.3 \\ 
\hline
  1500  & 641 & 2.65 & $1554 \pm 52$  & 17 \\ 
\hline
  2000  & 531 & 0.85 & $1940 \pm 80$  & 16 \\ 
\hline
  2500  & 1032 & 1.96 & $2391 \pm 84$  & 10 \\
\hline
\end{tabular}
\caption{The best kinematic cuts for representative $Z^\prime$ masses in the $Z^\prime$ leptophilic model. The transverse momentum and $e^+e^-$ mass are given in GeV. In the rightmost column, we display the signal efficiency, while the background efficiency is negligible in all cases.}
\label{TableII}
\end{table}

\begin{table}[t]
\centering
\begin{tabular}{|c|c|c|c|c|} 
\hline
${M}_{{Z}^{\prime}}$ & ${p}_{T}(l)>$ & $|\eta(l)|<$ & $|M_{ee}-M_c| < \delta_M $ & $\epsilon_S$(\%) \\ 
\hline
   500 & 189 & 2.00 & $485 \pm 65$ & 11.1 \\ 
\hline
  1000  & 387 & 1.69 & $1043 \pm 52$ & 37.4 \\ 
\hline
  1500  & 444 & 0.92 & $1461 \pm 88$ & 20.1 \\ 
\hline
  2000  & 777 & 0.86 & $1970 \pm 59$  & 30.4 \\ 
\hline
  2500  & 991 & 0.45 & $2547 \pm 64$  & 15 \\
\hline
\end{tabular}
\caption{The best kinematic cuts for representative $Z^\prime$ masses in the 331 models. The transverse momentum and $e^+e^-$ mass are given in GeV. In the rightmost column, we display the signal efficiency, while the background efficiency is negligible in all cases.}
\label{TableIII}
\end{table}

The best cuts for some $Z^\prime$ masses are shown in  Table~\ref{TableII} for the $Z^\prime$ leptophilic model and in Table~\ref{TableII} for the 331 models. The cut in the $e^+e^-$ mass was performed by searching for the best window around the signal peak to select the events. For each $Z^\prime$ mass, $4\times 10^5$ random searches were performed in the cut thresholds space of $p_T(l)$, $|\eta(l)|$, and $e^+e^-$ window around the signal peak. The search is agnostic of the $Z^\prime$ mass and the algorithm is able to identify the resonance peak without any previous information about the model parameters. \newline

The background efficiencies are tiny for each $Z^\prime$ mass, while the signal efficiency increases from light towards heavy $Z^\prime$ bosons, reaching a maximum of around 2.5 TeV masses in all models. From Table~\ref{TableII} and ~\ref{TableIII}, we see that higher signal efficiencies are achieved by hardening the $p_T$ threshold and selecting events that are more centrally produced in the detector. \newline

\begin{figure*}[!ht]
    \centering
         \subfigure[]{\includegraphics[scale=0.6
]{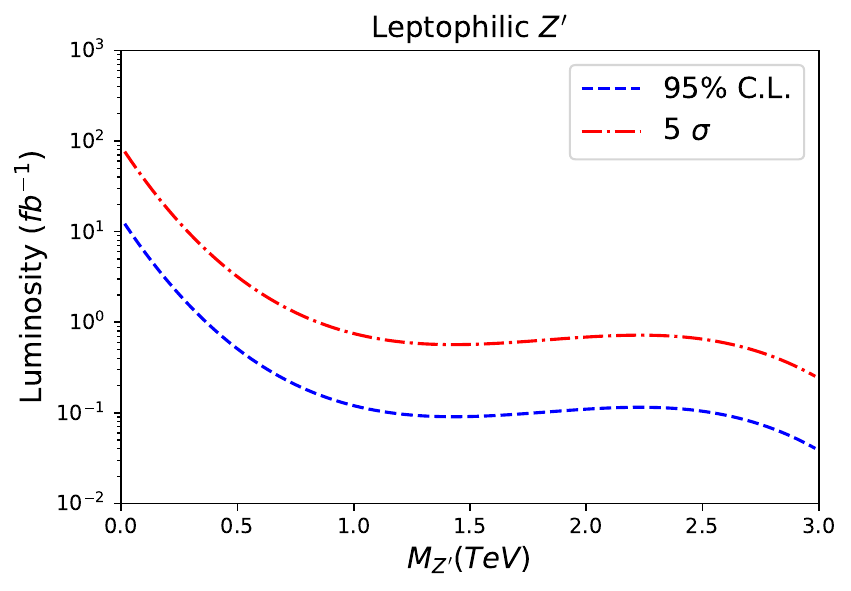}}
\subfigure[]{\includegraphics[scale=0.6]{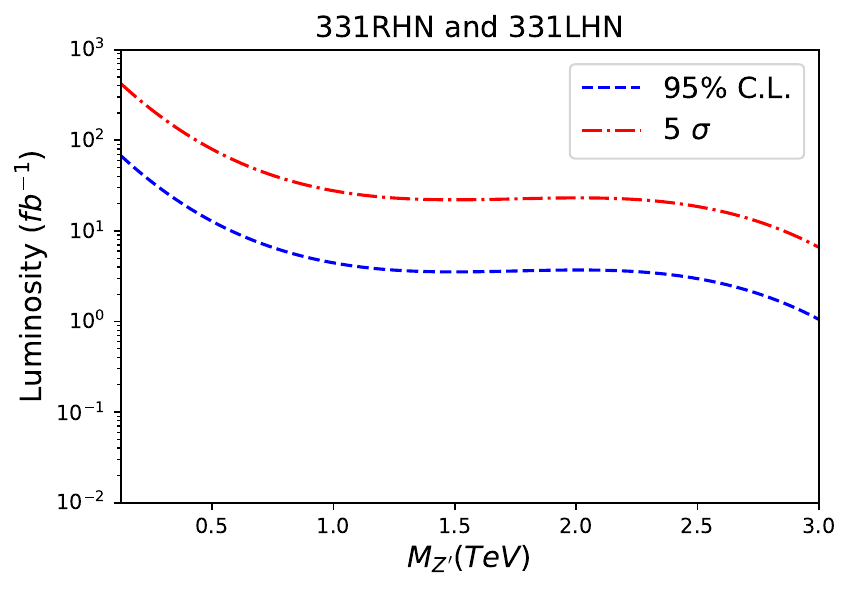}}
    \caption{The required luminosity, in fb$^{-1}$,  for 95\% CL exclusion and $5\sigma$ discovery, in 3 TeV CLIC, of the  leptophilic (left panel) and 331RHN/331LHN $Z^\prime$ boson (right panel).}
\label{331LHN_all_BMs}
\end{figure*}

We show, in Figure~\ref{331LHN_all_BMs}, the luminosity required to exclude a $Z^\prime$ at 95\% confidence level (CL) or to discover its signal in the 3 TeV CLIC. A 3-3-1 $Z^\prime$ will demand around one order of magnitude more data compared to leptophilic ones. Yet, 1 ab$^{-1}$ will suffice for the less promising case whilst luminosities as low as 100 pb$^{-1}$ will be needed for the most promising scenarios. It is known that LHC already places $M_{Z^\prime} > 4$~TeV \cite{Alves:2022hcp}. Therefore, CLIC would be represent a complementary search rather than a discovery one.  Nevertheless, it is worth pointing out that in the Multi-TeV $Z^\prime$ mass regime, CLIC has the potential to discover such a boson with less than $1fb^{-1}$ of data (See Figure~\ref{331LHN_all_BMs}b). This clearly shows that even if LHC happens to discover a $Z^\prime$ boson with a mass larger than CLIC's center of mass energy, CLIC will still play an important role in constraining its properties in a similar vein to LEP back at the time. \newline

Anyway, there is room for improvement if we also take final state muons into account which should nearly double the signal cross section after cuts depending on the $Z^\prime$ mass.  One can discriminate the leptophilic from the 3-3-1 hypothesis by looking for a resonance at different channels, such as dijets. \newline

An interesting possibility is that the $Z^\prime$ mediates the interactions between dark matter and the SM spectrum. In that case, due to the diminished branching ratio into leptons, the integrated luminosity to reach the sensitivity we forecast here will be larger but, given these excellent prospects for discovery and exclusion, observing this new boson at CLIC will probably remain viable. We plan to investigate the $Z^\prime$ phenomenology at CLIC in scenarios with a dark matter candidate in the future. \newline

\section{\label{SectionIV}CONCLUSIONS}
In this work, we investigate the prospects of the 3 TeV CLIC to unravel new physics associated with a new weak interaction coupling to electrons and positrons. If this new interaction couples feebly to quarks, then an $e^+e^-$ collider is the suitable machine to search for that leptophilic interaction. We found that an optimized kinematics cuts strategy search for $e^+e^-\to e^+e^-$ benefits from resonant and non-resonant contributions from a $Z^\prime$ improving the mass reach of the collider. \newline

In the leptophilic case, $Z^\prime$ masses from 100 GeV up to 3 TeV can be excluded at 95\% C.L. with less than $10 fb^{-1}$ if the strength of couplings with leptons is similar to the SM $Z$ boson, and with up to $100 fb^{-1}$ for a discovery. By their turn, LHN and RHN 3-3-1 models will need more data once couplings to quarks compete for the branching ratios of the $Z^\prime$. Yet, CLIC will surely complement searches performed at the LHC for 3-3-1 models.

In summary, we proposed that in the same way the sequential $Z^\prime$ boson is a benchmark model for LHC collaboration, a leptophilic $Z^\prime$ with sequential couplings to leptons be for the CLIC. We justified this argument by assessing the CLIC discovery potential, which reaches a $5\sigma$ detection with less than $1fb^{-1}$ of integrated luminosity for $M_{Z^\prime}=1-3$~TeV.

\acknowledgments

The authors thank Yoxara Villamixar for discussions. This work was supported by Simons Foundation (Award Number:1023171-RC), FAPESP Grant 2018/25225-9, 
2021/01089-1, 2023/01197-4, ICTP-SAIFR FAPESP Grants 2021/14335-0, CNPq Grants 307130/2021-5, 311851/2020-7, 307317/2021-8, and ANID-Millennium Science Initiative Program ICN2019\textunderscore044. Y.M.O.T. acknowledges financial support from CAPES under grants 88887.485509 / 2020-00. V.P.G. was partially supported by CNPq, CAPES, FAPERGS and  INCT-FNA (Process No. 464898/2014-5)

\nocite{*}


\bibliography{ref}
\end{document}